\documentclass[twocolumn,prl,secnumarabic,
amsmath,amssymb,superscriptaddress,nofootinbib]
{revtex4-2}

\usepackage{graphicx} 
\usepackage{xcolor}

\newcommand{\rf}[1]{Fig.~\ref{#1}} 

\newcommand{\ket}[1]{\left|#1\right>}
\newcommand{\bra}[1]{\left<#1\right|}

\newcommand{\f}[1]{\mbox{\boldmath$#1$}}
\newcommand{\fk}[1]{\mbox{\boldmath$\scriptstyle#1$}}

\newcommand{\bea}{\begin{eqnarray}}
\newcommand{\ea}{\end{eqnarray}}
\newcommand{\eea}{\end{eqnarray}}
\newcommand{\ord}{\,{\cal O}}

\begin{document}

\title{Axion-induced Casimir force between nuclei and dynamical axion 
pair creation} 

\author{Stefan Evans}
\email{s.evans@hzdr.de}
\affiliation{Helmholtz-Zentrum Dresden-Rossendorf, 
Bautzner Landstraße 400, 01328 Dresden, Germany}
\author{Ralf Sch\"utzhold}
\affiliation{Helmholtz-Zentrum Dresden-Rossendorf, 
Bautzner Landstraße 400, 01328 Dresden, Germany}
\affiliation{Institut f\"ur Theoretische Physik, 
Technische Universit\"at Dresden, 01062 Dresden, Germany}

\begin{abstract}
We study the interaction between axions and nuclei by combining the 
Peccei-Quinn mechanism with results from quantum chromo-dynamics (QCD) 
which imply that the QCD condensates are reduced within nuclear matter. 
Thus, the effective axion mass is also reduced, yielding a 
finite axion-nucleon scattering cross section.  
Even in the absence of real axions, this interaction would manifest itself 
in a Casimir type attraction between two nuclei. 
Finally, accelerated nuclei can create entangled pairs of axions via the 
dynamical Casimir effect (or as signatures of the Unruh effect).
\end{abstract}

\date{\today}

\maketitle

\paragraph{Introduction}

In the history of physics, open questions and problems have often motivated the 
prediction of new particles, which have later been observed experimentally.
Presently, axions are prominent candidates for such new particles. 
In the seminal work~\cite{Peccei:1977hh, Peccei:1977ur} by Peccei and Quinn, 
they were introduced in order 
to solve the strong $CP$ problem, i.e., the question of why quantum 
chromo-dynamics (QCD) is, at least to extremely good accuracy, invariant under 
charge $C$, parity $P$, and thus also time $T$ reversal 
(see below and~\cite{Weinberg:1977ma, Wilczek:1977pj, Kim:1979if, 
Shifman:1979if, Zhitnitsky:1980tq, Dine:1981rt, Peccei:2006as}). 
In addition, axions are also promising candidates for dark matter -- provided 
that they live long enough~\cite{Preskill:1982cy, Abbott:1982af, Dine:1982ah, 
Sikivie:1983ip, Sikivie:2006ni}. 

So, naturally, many schemes for axion searches have been proposed and 
realized, but so far without finding conclusive evidence for axions, see,
e.g.~\cite{ParticleDataGroup:2022pth}.
For example, the well studied anisotropies of the cosmic microwave background 
radiation~\cite{Cadamuro:2011fd} and astronomical data regarding the evolution 
of stars~\cite{Bernabei:2001ny, Raffelt:2006cw, Caputo:2024oqc, Ayala:2014pea, Irastorza:2018dyq, 
DeAngelis:2007dqd, Fortin:2021cog} or white dwarfs~\cite{Isern:2008nt, Dolan:2021rya} 
could be affected by axions and thus allow us to place restrictions on the 
possible axion parameter space, see also~\cite{Jaeckel:2010ni, Baker:2013zta, 
Buschmann:2019icd, CAST:2017uph, Dent:2020jhf}. 
Another class of axions searches are earth-based experiments.
Here, prominent examples include the ``light shining through wall'' 
experiments~\cite{Ruoso:1992ai, Cameron:1993mr, Fouche:2008jk, Redondo:2010dp, 
GammeVT-969:2007pci, Afanasev:2008jt, Ehret:2010mh, OSQAR:2015qdv, 
Inada:2016jzh, Beyer:2021mzq}, collider probes~\cite{Dobrich:2015jyk, Knapen:2016moh, Dolan:2017osp, 
Mariotti:2017vtv, FASER:2018eoc, NA64:2020qwq, Gori:2020xvq},
and ``light-by-light scattering'' 
schemes~\cite{Maiani:1986md, Raffelt:1987im, Semertzidis:1990qc, 
Ejlli:2020yhk, Battesti:2008, Fan:2017fnd, Evans:2023jpr}.

The involved energy scales determine which axion mass ranges these schemes 
are most sensitive to. 
In addition, the majority of these schemes do also depend crucially 
on the model-dependent 
coupling strengths between axions and other particles, such as photons or 
electrons -- which are also not known {\it a priori} -- or on additional 
assumptions, e.g., regarding the stellar evolution or the axion lifetime. 
In order to avoid this limitation, we restrict ourselves to a minimal, 
model-independent set of assumptions in the following. 
Regarding the axions, we just employ the Peccei-Quinn mechanism, 
which was the motivation for considering axions in the first place. 
Combining this idea with results from QCD, we derive the interaction 
between axions and nuclei. 
Thus, our results do only depend on the (hadronic contribution to the) axion mass, 
and not on other unknown coupling strengths. 

\paragraph{Strong CP problem}

Let us briefly review the basics. 
The field-strength tensor of the strong interaction is given by 
$G_{\mu\nu}^a=\partial_\mu A_\nu^a-\partial_\nu A_\mu^a+
f^a_{\;bc}A^b_\mu A^c_\nu$ in terms of the Yang-Mills fields $A_\mu^a$
and the $SU(3)$ structure constants $f^a_{\;bc}$ 
while the dual tensor reads 
$\tilde G^{\mu\nu a}=\epsilon^{\mu\nu\rho\sigma}G_{\rho\sigma}^a$. 
In analogy to the two Lorentz invariants $\f{E}^2-\f{B}^2$ and  
$\f{E}\cdot\f{B}$ in electrodynamics, we may form the two contractions 
$G_{\mu\nu}^a G^{\mu\nu}_a$ and $G_{\mu\nu}^a\tilde G^{\mu\nu}_a$
leading to the Lagrangian ($\hbar=c=1$)
\bea
\label{qcd-theta} 
{\cal L}_{\rm gluons}
=
-\frac{1}{4g^2_{\rm QCD}}\,G_{\mu\nu}^a G^{\mu\nu}_a 
+\frac{\theta}{32\pi^2}\,G_{\mu\nu}^a\tilde G^{\mu\nu}_a 
\,,
\ea
where $g_{\rm QCD}$ is the QCD coupling strength~\cite{Schafer:1996wv}.
Again in analogy to electrodynamics, the first term 
$G_{\mu\nu}^a G^{\mu\nu}_a$ is even under charge $C$, parity $P$, 
and time $T$ reversal, while the second term 
$G_{\mu\nu}^a\tilde G^{\mu\nu}_a$ is odd under $CP$ and $T$. 

For $\theta=0$, the theory is $CP$ invariant and thus expectation
values of $CP$ odd quantities such as $G_{\mu\nu}^a\tilde G^{\mu\nu}_a$ 
vanish. 
Allowing for a small but finite $\theta$, however, this is no longer true. 
As already expected from stationary perturbation theory 
(treating $\theta$ as the small expansion parameter), 
these quantities may now acquire non-vanishing expectation values 
which scale linearly in $\theta$ for small $\theta$. 
Specifically, one finds (to leading order)  
\bea
\label{quark}
\langle\hat G_{\mu\nu}^a\hat{\tilde G}^{\mu\nu}_a\rangle
=
32\pi^2\langle\hat{\bar\Psi}\hat\Psi\rangle 
\frac{m_um_d}{m_u+m_d}\,\theta
\,,
\ea
where $m_u$ and $m_d$ are the masses of the up and down quarks 
while $\langle\hat{\bar\Psi}\hat\Psi\rangle\approx-10^{-2}\rm GeV^3$ 
is the quark condensate, see, e.g.~\cite{Shifman:1979if}. 

In a similar way, the $\theta$-term would also generate other 
$CP$-violating expectation values and observables, which are not 
observed experimentally. 
Present experimental bounds (e.g., from the electric dipole moment 
of the neutron~\cite{Abel:2020pzs}) limit this parameter to $|\theta|<10^{-9}$. 
The question of why this parameter is so small (or even zero) 
is usually referred to as the strong $CP$ problem. 

\paragraph{Peccei-Quinn mechanism}

In order to solve this strong $CP$ problem, the idea is to postulate a 
massless pseudo-scalar axion field $\phi$ coupled to the 
$G_{\mu\nu}^a\tilde G^{\mu\nu}_a$ 
term 
\bea
\label{axion}
{\cal L}_{\rm axion}
=\frac12 (\partial_\mu\phi) (\partial^\mu\phi) 
+
\frac{g_\phi}{32\pi^2} G_{\mu\nu}^a\tilde G^{\mu\nu}_a \phi 
\,.
\ea
Since $\phi$ is massless, the $\theta$-term in Eq.~\eqref{qcd-theta}
can be absorbed by a simple re-definition of 
$\phi\to\phi-\theta/g_\phi$. 
After that, the equation of motion 
$\Box\phi=g_\phi G_{\mu\nu}^a\tilde G^{\mu\nu}_a/(32\pi^2)$ 
can be combined with Eq.~\eqref{quark} to show that the 
expectation value $\phi=\langle\hat\phi\rangle$ behaves as a massive 
scalar field where the effective axion mass $m_\phi$ is determined by 
(see also \cite{Balkin:2020dsr})
\bea
\label{axion-mass}
m_\phi^2
=
-g_\phi^2
\langle\hat{\bar\Psi}\hat\Psi\rangle 
\frac{m_um_d}{m_u+m_d}
\,.
\ea
Note that the quark condensate $\langle\hat{\bar\Psi}\hat\Psi\rangle$
is negative.

Even though this line of arguments is extremely simplified and leaves 
out many important details, it allows us to grasp the main idea.
In analogy to a superconductor where the coupling between the 
electromagnetic field $A_\mu$ and the currents $j^\mu$ inside the 
superconductor generate an effective mass for the photons, the QCD 
fluctuations coupled to the axion field generate an effective mass 
$m_\phi$ term for the latter.
Thus, in the ground state, the electromagnetic field vanishes in 
the superconductor (Mei\ss ner-Ochsenfeld effect). 
Similarly, the $CP$-violating term vanishes in vacuum
$\phi=\langle\hat\phi\rangle=0$, which is 
how the Peccei-Quinn mechanism solves the strong $CP$ problem. 

\paragraph{Nuclear matter}

After having sketched the Peccei-Quinn mechanism in vacuum, 
let us consider the situation in nuclear matter, say inside a large 
nucleus (for the interior of neutron stars, see, 
e.g.,~\cite{Hook:2017psm, Balkin:2020dsr, Balkin:2022qer, DiLuzio:2021pxd, 
Anzuini:2023whm}). 
It is by now well established that the QCD fluctuations such as the 
quark condensate $\langle\hat{\bar\Psi}\hat\Psi\rangle$ are reduced 
inside nuclei. 
In very simplified models, such as the most simple bag model where 
the nuclear interior is modeled as a free field, these condensates 
are even reduced to zero. 
More sophisticated models yield a reduction between 
25 and 50 percent~\cite{Cohen:1991nk, Balkin:2020dsr}.

Retracing the line of arguments above, we see that a reduction of the 
quark condensate $\langle\hat{\bar\Psi}\hat\Psi\rangle$ inside the nuclei 
would also imply a reduction of the effective axion mass, i.e.,  
$m_\phi^2\to m_\phi^2[1-\varepsilon]$ where we have denoted the reduction 
ratio by $\varepsilon$. 
Employing the aforementioned analogy to superconductors, a reduction of 
the superfluid density (e.g., by defects or a finite temperature) would 
lead to a suppression of the effective photon mass and thus a growth of 
the London penetration depth. 

In view of the above arguments, the reduction of the axion mass 
(or, more precisely, the curvature of the axion potential) is a fairly 
robust result and largely independent of other model parameters and 
assumptions.
Even though there is a slight uncertainty regarding the precise value of 
$\varepsilon$, the situation is quite different from other axion searches, 
where the considered range for the axion-photon interaction strength, 
for example, can span several orders of magnitude. 

Note that the coupling between axions and gluons in Eq.~\eqref{axion} 
suggests, via Feynman diagrams containing internal quark loops, for example,
the possibility of an effective coupling between axions and photons. 
However, there are several potential reasons why this effective coupling
could be suppressed (which might also explain why we did not see axions yet). 
As one option, this effective coupling could be (partially) canceled by 
another contribution of different origin. 
As another option, interaction effects such as running coupling or confinement 
could suppress this effective coupling
\cite{Jain:2005nh, Masso:2005ym, Masso:2006gc, Jaeckel:2006xm}. 
For example, the axion field could actually be effectively coupled to a 
suitable spatio-temporal average of $G_{\mu\nu}^a\tilde G^{\mu\nu}_a$ 
such that the coupling strength $g_\phi$ is most effective at low 
momentum scales. 
Together with quark confinement, this could suppress the 
Feynman diagrams containing quark loops.

\paragraph{Scattering cross section}

Incorporating the reduction of the effective axion mass inside the nucleus, 
the axion evolution equation becomes 
\bea
\label{Klein-Fock-Gordon}
\left(\Box+m_\phi^2\right)\phi=m_\phi^2\varepsilon(\f{r})\phi 
\,.
\ea
where $\varepsilon(\f{r})$ vanishes in vacuum and describes the 
reduction inside the nucleus. 
In order to ensure the validity of this effective description, 
we assume that the energy and momentum scales of the axion field $\phi$ 
are much smaller than the characteristic scales of the QCD fluctuations,
which are governed by $\Lambda_{\rm QCD}=\ord(100~\rm MeV)$. 

For a fixed axion frequency $\partial_t\phi\to-i\omega\phi$, 
the above Klein-Fock-Gordon equation assumes the same form as 
a stationary Schr\"odinger equation describing scattering at a 
spherical well potential.
Thus, we may apply the same methods and read off the differential 
scattering cross section $d\sigma/d\Omega$. 
For axion wavelengths much larger than the nuclear radius, 
the nucleus acts as a delta potential and thus we find 
(to lowest order in $\varepsilon$, 
see also~\cite{Fukuda:2021drn,VanTilburg:2024tst}) 
\bea
\label{cross-section}
\frac{d\sigma}{d\Omega}= 
\frac{ m_\phi^4\varepsilon^2V_{\rm nucl}^2 }{16\pi^2} 
\,,
\ea
where $V_{\rm nucl}$ denotes the volume of the nucleus. 
E.g., for $\varepsilon=0.3$~\cite{Balkin:2020dsr} and 
an axion mass of $m_\phi=10~\rm MeV$, we find a cross section of 
$\sigma=40~\mu\rm b$ for an iron nucleus.  
To visualize this quantity, it would correspond to a classical mean 
free path of 250~m for the axion in solid iron. 

If the axion wavelength is much longer than the distance between the 
iron nuclei, this picture of a classical mean free path would no longer 
apply and the iron would instead act as an effective medium.
Thus an iron cylinder could act like an optical fiber for very slow 
axions, or one could imagine a magnifying lens for axions, for example. 

If, on the other hand, the axion wavelength becomes comparable to the 
nuclear radius, we get the well-known resonance effects. 
For even shorter axion wavelengths, the precise internal structure 
of the nuclei, i.e., how $\varepsilon(\f{r})$ is switched on and off, 
starts to play a role -- but at that point, one should also be careful 
and consider the region of validity of the above effective 
description~\eqref{Klein-Fock-Gordon}. 

\paragraph{Casimir attraction between nuclei}

Even in the absence of real axions, their inevitable quantum fluctuations 
can also induce potentially measurable effects. 
As one example, let us study the induced Casimir attraction between two 
nuclei.
To this end, let us start from the perturbation Hamiltonian of the 
quantized axion field $\hat\phi$ 
\bea
\label{perturbation}
\hat H_{\rm int}=\frac{m_\phi^2}{2}\int d^3r\,\varepsilon(\f{r})\hat\phi^2 
\,.
\ea
Using $\varepsilon$ as a small expansion parameter, we may apply 
second-order stationary perturbation theory to obtain the ground-state 
energy shift 
\bea
\label{second-order}
\Delta E_{(2)}=\sum\limits_{\ket{\Psi}\neq\ket{0}}
\frac{|\bra{\Psi} \hat H_{\rm int} \ket{0}|^2}
{E_{\ket{\Psi}}-E_{\ket{0}}}
\,.
\ea
In view of the structure of the perturbation Hamiltonian~\eqref{perturbation},
we see that two-axion states $\ket{\Psi}=\ket{\f{k}_1,\f{k}_2}$ are the 
relevant intermediate states. 

Let us first consider the non-relativistic limit where the two axion 
momenta $\f{k}_1$ and $\f{k}_2$ are well below the axion mass $m_\phi$.
Then we may approximate the energy denominator in Eq.~\eqref{second-order} by 
${E_{\ket{\Psi}}-E_{\ket{0}}}\approx2m_\phi$ which yields the energy shift 
$\Delta E_{(2)}\approx\bra{0}(:\hat H_{\rm int}:)^2\ket{0}/(2m_\phi)$ 
containing the perturbation Hamiltonian~\eqref{perturbation} in normal 
ordering. 
Using the Wick theorem and focusing on those contributions which depend on 
the distance between the nuclei, which give rise to the Casimir effect,
we find 
\bea
\Delta E_{\rm Casimir}
\approx 
\frac{m_\phi^3\varepsilon^2V_{\rm nucl}^2}{8}
(\bra{0}\hat\phi(\f{r}_1)\hat\phi(\f{r}_2)\ket{0})^2
\,,
\ea
where we have assumed that the nuclear radii are much smaller than 
all other relevant length scales such that we may approximate the 
nuclei as point particles at the positions $\f{r}_1$ and $\f{r}_2$. 

Here $\bra{0}\hat\phi(\f{r}_1)\hat\phi(\f{r}_2)\ket{0}=W(\ell)$ 
denotes the usual two-point function 
$W(\ell)=m_\phi K_1(m_\phi\ell)/(4\pi^2\ell)$ 
of a massive scalar field depending on the distance $\ell=|\f{r}_1-\f{r}_2|$ 
where $K_1$ is the modified Bessel function of the second kind.
At large distances $m_\phi\ell\gg1$ (which is the case considered above), 
it obeys the asymptotic form 
$W(\ell)\sim\sqrt{m_\phi/(32\pi^3\ell^{3})}\,e^{-m_\phi\ell}$, see, e.g., 
\cite{Bogolyubov:1959bfo}.
 
In view of the exponential suppression at large separations $\ell$, 
the Casimir force becomes most pronounced at smaller distances.
In this regime, however, the approximation 
${E_{\ket{\Psi}}-E_{\ket{0}}}\approx2m_\phi$
is no longer applicable and we have to perform the sum over all 
intermediate two-axion states $\ket{\Psi}=\ket{\f{k}_1,\f{k}_2}$,
i.e., the $d^3k_1$ and $d^3k_2$ integrations. 
For very small distances, the precise functional form of 
$\varepsilon(\f{r})$ starts to play a role, which depends on 
the nuclear model. 
We modeled $\varepsilon(\f{r})$ by a Gaussian for simplicity. 
For two iron nuclei, the results are shown in 
\rf{casimirFigIron} as a function of axion mass $m_\phi$ and distance 
$\ell=|\f{r}_1-\f{r}_2|$.  
E.g., for $m_\phi=50$~MeV and rather small distances of order 10~fm 
(i.e., just above twice the nuclear radius), we obtain a Casimir energy of order
10~eV.
Going away from a Gaussian profile $\varepsilon(\f{r})$ and 
considering sharper (i.e., more step-like) switching functions, 
one would expect to obtain larger Casimir forces.

%
\begin{figure}[h]
\center
\includegraphics[width=0.9\columnwidth]{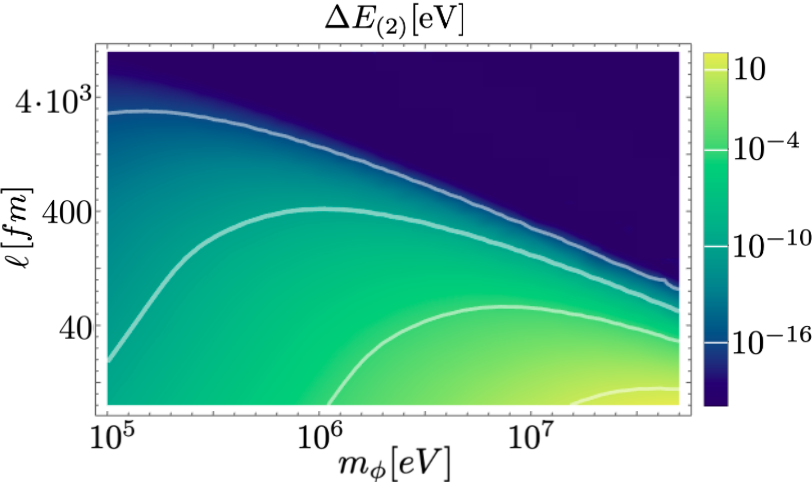}
\caption{\label{casimirFigIron}
Casimir energy between two iron nuclei, as a function of axion mass and distance.} 
\end{figure}
%
%

Of course, the strong Coulomb repulsion between the iron nuclei 
of order $100$~MeV would render the measurement of such a small 
attractive force quite difficult. 
For smaller axion masses, which would lead to Casimir forces of longer range, 
one could consider iron atoms instead, where the Coulomb repulsion is 
shielded by the electrons. 
This would be very similar to ``fifth-force'' measurements, see, 
e.g., \cite{VanTilburg:2024tst, Capolupo:2021dnl, Heacock:2021btd, 
Bauer:2023czj, Barbosa:2024tty, Kim:2022ype, Banerjee:2022sqg}. 

As another option, we could push our analysis to the extreme limit and 
consider two protons or neutrons, together 
with slightly heavier axions up to 100~MeV mass.  
For very small distances between 2 and 3~fm, the Casimir energy would be 
of the same order, between $\ord(10~\rm eV)$ and $\ord(100~\rm eV)$.
However, the Coulomb repulsion between two protons would be much smaller 
than in the iron case above, such that the Casimir energy might amount to 
a correction in the per-cent to per-mill range.  
For two neutrons, the dominant force is the magnetic dipole-dipole interaction.
Here the Casimir energy may induce a correction 
also in the per-cent to per-mill range, 
provided that the dipoles are aligned. 
For random dipole orientations, the dipole-dipole interactions would average 
out such that only the Casimir effect remains.

Note that our scattering potential applied to neutrons is 
similar to the quadratic axion-nucleon coupling considered previously 
\cite{Fukuda:2021drn, Bauer:2023czj, VanTilburg:2024tst} 
which for light axions has astrophysical and fifth force implications. 

\paragraph{Accelerated nuclei}

After having discussed the static Casimir effect, let us consider a dynamical 
scenario involving accelerated nuclei. 
One way of realizing such a scenario would be to have ultra-peripheral 
collisions of nuclei in analogy to the famous Rutherford experiment.
In this case, the accelerated nuclei can create entangled axions pairs out of 
the vacuum, which is often referred to as the dynamical Casimir effect,
see, e.g.,~\cite{Moore:1970tmc, Davies:1976hi, Davies:1977yv, Dodonov:2020eto}. 
The amplitude for this process can be calculated with the same perturbation 
Hamiltonian~\eqref{perturbation} but now adapted to the case of moving nuclei.
To lowest order in $\varepsilon$, the amplitudes stemming from the two 
(or more) nuclei would just add up, so it is sufficient to consider one 
accelerated nucleus. 
Actually, using the same arguments as in~\cite{Schutzhold:2006gj}, 
the axion radiation emitted by such an accelerated nucleus can also be 
interpreted as a signature of the Unruh effect, 
see also~\cite{Fulling:1972md, Davies:1974th, Unruh:1976db, Schutzhold:2011ze}. 
Assuming a constant acceleration $a$ for simplicity, a co-moving observer 
would perceive the vacuum as a thermal bath of axions with the Unruh 
temperature $T_{\rm Unruh}=a/(2\pi k_{\rm B})$.
Thus, the nucleus can scatter an axion out of this thermal bath according 
to the cross section~\eqref{cross-section}. 
Translated back to the laboratory frame, this scattering process would 
correspond to the emission of an entangled pair of axions, 
see also~\cite{Schutzhold:2008zza}. 

In order to calculate the axion pair-creation amplitude, let us again assume 
that the nuclear radius is much smaller than all other relevant length scales
such as the axion wavelengths. 
Then we may approximate the nucleus as a point particle with the trajectory 
$\f{r}_{\rm nucl}(t)$ and derive the axion pair-creation amplitude via 
time-dependent perturbation theory with respect to $\hat H_{\rm int}$ 
\bea
\label{amplitude}
{\mathfrak A}_{\fk{k}_1\fk{k}_2}
=
\frac{\varepsilon m_\phi^2 V_{\rm nucl}}{2\sqrt{\omega_1\omega_2}}
\int dt\,e^{i(\omega_1+\omega_2)t-i(\fk{k}_1+\fk{k}_2)\cdot\fk{r}_{\rm nucl}(t)}
.\quad 
\ea
Following~\cite{Schutzhold:2006gj}, let us introduce the new time coordinate  
$\tau=t-(\f{k}_1+\f{k}_2)\cdot\f{r}_{\rm nucl}(t)/[\omega_1+\omega_2]$.
Assuming non-relativistic velocities $|\dot{\f{r}}_{\rm nucl}|\ll1$, 
we may simplify the amplitude in Eq.~\eqref{amplitude} to 
\bea
\label{amplitude-approx}
{\mathfrak A}_{\fk{k}_1\fk{k}_2}
\approx 
\frac{\varepsilon m_\phi^2 V_{\rm nucl}}{2\sqrt{\omega_1\omega_2}}
\,\frac{\f{k}_1+\f{k}_2}{\omega_1+\omega_2}\cdot
\tilde{\f{v}}_{\rm nucl}(\omega_1+\omega_2)
\,,
\ea
where $\tilde{\f{v}}_{\rm nucl}$ denotes the Fourier transform of 
$\dot{\f{r}}_{\rm nucl}(\tau)$. 
As a consequence, the axion pair-creation amplitude becomes maximal if 
both axions are emitted into the same direction $\f{k}_1\|\f{k}_2$. 

For the ultra-peripheral collision scenario, $\dot{\f{r}}_{\rm nucl}(\tau)$ 
is constant asymptotically and changes during a characteristic acceleration 
time scale $\Delta\tau$.
Thus, for small frequencies $\omega\ll1/\Delta\tau$, the Fourier transform 
$\tilde{\f{v}}_{\rm nucl}$ is basically that of a Heaviside step function 
$\tilde{\f{v}}_{\rm nucl}(\omega)\sim1/\omega$. 
For high frequencies $\omega\gg1/\Delta\tau$, on the other hand, the 
Fourier transform $\tilde{\f{v}}_{\rm nucl}$ falls off very rapidly since 
$\dot{\f{r}}_{\rm nucl}(\tau)$ is a smooth function. 

The total axion pair-creation probability $\mathfrak P$ is obtained by 
integrating $|{\mathfrak A}_{\fk{k}_1\fk{k}_2}|^2$ over the two momenta 
$\f{k}_1$ and $\f{k}_2$. 
Its scaling depends on the axion mass $m_\phi$ in comparison to the 
acceleration time scale $\Delta\tau$.
For large axion masses $m_\phi\gg1/\Delta\tau$, the pre-factor $m_\phi$
in Eq.~\eqref{amplitude-approx} would be large, but the Fourier transform 
$\tilde{\f{v}}_{\rm nucl}(\omega_1+\omega_2)$ would be strongly suppressed. 
Thus, a larger pair-creation probability $\mathfrak P$ is obtained
for small axion masses $m_\phi\ll1/\Delta\tau$. 
In this case, the created axions are predominantly ultra-relativistic 
and thus we may approximate $\omega_1\approx|\f{k}_1|$ and 
$\omega_2\approx|\f{k}_2|$. 
Integrating $|{\mathfrak A}_{\fk{k}_1\fk{k}_2}|^2$ up to the cut-off 
scale $\omega_{\rm cut}\approx|\f{k}_{\rm cut}|\sim1/\Delta\tau$, 
we find the scaling 
\bea
\label{scaling}
{\mathfrak P}=\ord\left(\frac{\varepsilon^2 m_\phi^4 V_{\rm nucl}^2}
{(\Delta\tau)^2}\right) 
\,.
\ea
In view of the assumed hierarchy $V_{\rm nucl}^{1/3}\ll\Delta\tau\ll1/m_\phi$, 
this is a small number, but considering many collisions, the total probability 
might become non-negligible. 
Furthermore, by pushing our analysis to the limit where 
$V_{\rm nucl}^{1/3}\sim\Delta\tau\sim1/m_\phi$, we would expect 
${\mathfrak P}=\ord(\varepsilon^2)$, but the precise value requires 
more careful calculations because the approximations we used break down. 
Of course, measuring the created axion pairs is another matter, 
but this also applies to other axion searches.

\paragraph{Conclusions and Outlook}

We combine the Peccei-Quinn mechanism with results from QCD.
Since the quark condensate changes inside nuclear matter,  
we deduce an analogous shift of the effective axion mass. 
In this way, we obtain an effective axion interaction with nuclei or 
other objects that generate inhomogeneities in the condensate. 
This description is quite robust and largely independent of other parameters 
(such as the axion-photon coupling strengths) or model assumptions.
Note that we only deduce a change of the axion mass, i.e., the curvature of 
the axion potential -- the expectation value of the axion field is not modified 
(in contrast to previous works~\cite{Hook:2017psm, Balkin:2020dsr, 
Balkin:2022qer, DiLuzio:2021pxd, Anzuini:2023whm} 
which consider a change of the expectation value inside neutron stars, 
for example).

As our first application, we derive scattering cross sections $\sigma$ 
between axions and nuclei, which only depend on the axion mass $m_\phi$
and the nuclear radius. 
E.g., for an iron nucleus and $m_\phi=10~\rm MeV$, 
we find $\sigma=40~\mu\rm b$. 
These results should be taken into account when analyzing axion bounds 
from astronomical data. 
For example, considering the energy loss of a white dwarf~\cite{Dolan:2021rya} 
with a radius of $10^7~\rm m$, composed of oxygen with a density of $10^5$ 
of that of ordinary matter, axions in the mass range $m_\phi\geq100~\rm keV$ 
would have a mean free path of $\leq10^6~\rm m$ and thus be effectively trapped.
Similar arguments apply to other dense stellar media, e.g., in supernovae  
\cite{Turner:1987by, Lucente:2022vuo, Chang:2018rso, Ertas:2020xcc, 
Calore:2021klc, Dunsky:2022uoq}, as well as for earth-based axion search schemes. 
For very low-energy axions, one might even imagine interesting possibilities 
such as axion wave-guides or magnifying lenses for axions.

Even in the absence of real axions, their inevitable quantum vacuum
fluctuations would generate a Casimir type attraction between nuclei,
where the range of the force is governed by the axion mass. 
For the parameters considered here (see \rf{casimirFigIron}),
the Casimir energies lie well below the Coulomb repulsion, 
but speculating about larger values of $\varepsilon m_\phi^2$ 
and depending on the precise shape of the nuclei, this situation 
might change and the Casimir contribution could perhaps become 
non-negligible for understanding nuclear interactions, e.g., 
the deuteron binding energy of around 2~MeV.

Axion production is also possible, depending on the space-time dependence 
of the quark condensate. 
We show that in ultra-peripheral heavy ion collisions, the acceleration of 
nuclei, i.e., of the dips in the quark condensate, may generate entangled axion 
pairs akin to the dynamical Casimir effect or as signatures of 
Unruh radiation~\cite{Schutzhold:2006gj}. 
As an outlook, ultra-strong magnetic fields generated by such 
ultra-peripheral collisions \cite{Ruffini:2009hg, Grayson:2022asf} 
may change the quark condensate by a few percent too 
\cite{Shushpanov:1997sf, Klevansky:1989vi, Suganuma:1990nn, DElia:2011koc, 
Watson:2013ghq}, 
a complement to the recently 
developed quadratic photon-axion coupling~\cite{Kim:2023pvt, Beadle:2023flm}, 
which can also lead to the creation of
axion pairs with a scaling analogous to Eq.~\eqref{scaling}.


\end{document}